\documentclass[prl,10pt,aps,floatfix,notitlepage,twocolumn,superscriptaddress]{revtex4-2}

\usepackage{aas_macros}
\usepackage{amsmath,amsfonts,amssymb}
\usepackage[colorlinks=true,citecolor=blue,urlcolor=blue]{hyperref}
\usepackage{graphicx}
\usepackage{mathrsfs}
\usepackage{xcolor}
\usepackage{mathtools, bm}
\usepackage{amsmath}
\usepackage{braket}
\usepackage{physics}
\usepackage{siunitx}
\usepackage{booktabs}
\usepackage{microtype} 
\usepackage{comment}
\usepackage{soul}

\makeatletter
\renewcommand\onecolumngrid{%
  \do@columngrid{one}{\@ne}%
  \def\set@footnotewidth{\onecolumngrid}%
  \def\footnoterule{\kern-6pt\hrule width 1.5in\kern6pt}%
}
\renewcommand\twocolumngrid{%
  \def\footnoterule{%
    \dimen@\skip\footins\divide\dimen@\thr@@
    \kern-\dimen@\hrule width.5in\kern\dimen@}%
  \do@columngrid{mlt}{\tw@}%
}
\makeatother

\usepackage{accents}
\newlength{\dhatheight}

\makeatletter
\def\bstctlcite{\@ifnextchar[{\@bstctlcite}{\@bstctlcite[@auxout]}}
\def\@bstctlcite[#1]#2{\@bsphack
	\@for\@citeb:=#2\do{%
		\edef\@citeb{\expandafter\@firstofone\@citeb}%
		\if@filesw\immediate\write\csname #1\endcsname{\string\citation{\@citeb}}\fi}%
	\@esphack}
\makeatother

\begin{document}
	
\title{From scalar clouds around evaporating black holes to boson stars}

\author{Daniel Neves}
\email{daniel.neves@uc.pt}
\affiliation{Univ Coimbra, Faculdade de Ci\^encias e Tecnologia da Universidade de Coimbra and CFisUC, Rua Larga, 3004-516 Coimbra, Portugal}

\author{Jo\~{a}o G.~Rosa}
\email{jgrosa@uc.pt}
\affiliation{Univ Coimbra, Faculdade de Ci\^encias e Tecnologia da Universidade de Coimbra and CFisUC, Rua Larga, 3004-516 Coimbra, Portugal}

\date{\today}

\begin{abstract}
We study, for the first time, the evolution of a scalar cloud bound to an evaporating black hole. Our simulations of the associated Schr\"odinger-Poisson system for non-relativistic and spherically symmetric clouds reveal that a scalar cloud may (partially) survive as a self-gravitating boson star if the black hole evaporates adiabatically until its mass becomes less than one half of the cloud's mass. This yields a novel mechanism for boson star formation and shows that, as previously conjectured, bosonic dark matter production by light primordial black holes may result in micro-boson stars with very large occupation numbers, greatly enhancing their potential detectability even for very weakly interacting dark matter particles.
\end{abstract}

\maketitle

Matter fields can be bound to black holes (BHs) in a discrete Hydrogen-like spectrum  \cite{Damour:1976kh, Detweiler:1980uk, Cardoso:2004nk, Dolan:2007mj, Brito:2015oca}:
\begin{equation} \label{H_spectrum}
 \omega_n\simeq \mu -{\alpha^2\over 2n^2}\mu~,   
\end{equation}
where $n=1,2,\ldots$ is the principal quantum number. This approximation holds for small values of the dimensionless mass coupling $\alpha=\mu M_{BH}/M_P^2\ll 1$, which plays the role of the fine structure constant in atomic physics, where $\mu$ and $M_{BH}$ denote the matter field and BH masses, respectively, and $M_P=G^{-1/2}$ is the Planck mass (in natural units). This ``gravitational atom'' is simply a consequence of the Coulomb-like form of the BH's gravitational potential far away from the horizon, noting that the ``gravitational Bohr radius'' that sets the spatial extension of the bound state wavefunctions $a_0=(\alpha\mu)^{-1}=GM_{BH}/\alpha^2 \gg GM_{BH}$. The average velocities of matter particles are also limited by $\alpha$, which are thus non-relativistic for $\alpha\ll1$.

For bosonic matter fields these bound states admit large occupation numbers, which may be attained through different mechanisms. In the case of spinning BHs, the superradiant instability can populate states with non-zero angular momentum at the expense of the BH's rotational energy, particularly the ``2p-state'' with quantum numbers $(nlm)=(211)$. The number of particles grows exponentially fast within the superradiant regime $\omega_n< m\Omega_{BH}$, where $\Omega_{BH}$ denotes the BH's angular velocity, until this condition is saturated, yielding a total number $N_{211}\sim \tilde{a}(M_{BH}/M_P)^2\gg1$, where $0<\tilde{a}<1$ is the dimensionless BH spin parameter. 
The phenomenology and astrophysical relevance of the resulting scalar clouds (spin-0) and Proca clouds (spin-1) have been the subject of several studies in the recent literature \cite{Arvanitaki:2009fg, Rosa:2009ei, Arvanitaki:2010sy, Rosa:2011my, Pani:2012bp, Rosa:2012uz, Witek:2012tr, Brito:2013wya, Brito:2014wla, Arvanitaki:2016qwi,  Brito:2017wnc, Baryakhtar:2017ngi, Rosa:2017ury, Baumann:2018vus,   Ikeda:2018nhb, Berti:2019wnn, Baumann:2019eav, Cardoso:2018tly, Dolan:2018dqv, Edwards:2019tzf, Mehta:2020kwu,Blas:2020nbs, Ferraz:2020zgi, Baryakhtar:2020gao, Yuan:2021ebu, Caputo:2021efm, Tong:2022bbl, Ribeiro:2022ohq, Siemonsen:2022ivj, Spieksma:2023vwl, Branco:2023frw, Calza:2023rjt, Neves:2024rju}.

Stimulated capture of ambient dark matter particles, e.g.~through self-scattering processes \cite{Budker:2023sex}, has also been shown to be an efficient mechanism to populate bosonic BH bound states, in this case the (100) ground state \cite{Branco:2025znp}.

While for stellar-mass and supermassive BHs these processes generically result in effectively stable ``hairy-BHs'' \cite{Herdeiro:2014goa, Herdeiro:2016tmi}, for light primordial BHs (PBHs), potentially formed in the early Universe (see e.g.~\cite{Escriva:2022duf} for a recent review), the fate of scalar/Proca clouds is not yet understood, given that PBHs with $M_{BH}\lesssim 10^{12}$ kg evaporate away before the present day, at least within the semi-classical description of Hawking emission.

In \cite{March-Russell:2022zll} the superradiant instability was shown to be an efficient mechanism of producing heavy ($\gtrsim $TeV) dark matter bosonic particles (both spin-1 and spin-0), for light PBHs with $M_{BH}\lesssim 10^6$ kg that evaporate in less than a second, i.e.~before cosmological nucleosynthesis takes place (see also \cite{Bernal:2022oha}). This study identified parametric regimes where a dark scalar/Proca cloud can attain its maximum particle number before Hawking emission significantly depletes the PBH mass and spin. The decrease in the BH's angular velocity also makes the cloud exit the superradiant regime and thus be (partially) reabsorbed by the PBH. Since the reabsorption rate scales as $\alpha^9$, it quickly becomes negligible as the PBH mass decreases. 

In the late stages of PBH evaporation we may therefore have a scalar/Proca cloud with a large and essentially fixed particle number that become less and less bound to the PBH as it evaporates. Does the cloud simply dissipate away or can it remain, at least partially, self-bound?

Following standard arguments in quantum mechanical systems with time-varying potentials, \cite{March-Russell:2022zll} conjectured that at least part of the scalar/Proca clouds could survive as self-gravitating ``boson stars'' (of sub-atomic sizes) as long as evaporation occurs adiabatically until the PBH mass becomes less than the cloud's mass (generically smaller than the initial BH mass).

We may recast the discussion in \cite{March-Russell:2022zll} as follows. The BH mass decreases as $\dot{M}_{BH}\propto -M_{BH}^{-2}$ through Hawking emission. This implies that, initially, the BH's gravitational potential may change slowly, on time-scales longer than the bound state's oscillation period, and the wavefunction adjusts in an adiabatic fashion to the changing $\alpha$ coupling. The cloud essentially expands due to the increasing gravitational Bohr radius, although its shape also becomes more and more affected by its self-gravitational potential. 

Evaporation eventually becomes too fast for the wavefunction to adjust adiabatically, leading to what is commonly referred to as a ``quantum quench''. From the bound state's perspective, the BH simply vanishes, thus removing a fraction of the potential energy binding the cloud. For simplicity, assume that the cloud's self-gravitational potential is also Coulomb-like, so that the virial relation $\langle V\rangle = -2\langle K\rangle$ between the potential and the kinetic energy holds. It is then easy to conclude that, after the quench, the wave-function can only remain bound (with negative energy) if the BH mass is already smaller than the cloud's mass at the quench.

However, the cloud's self-gravity is only asymptotically Coulomb-like and the problem is, of course, non-linear, which means that a numerical study must be performed to verify the validity of the above qualitative arguments, as well as to accurately determine the conditions under which a scalar/Proca cloud can survive the BH's evaporation in a self-gravitating configuration.

In this Letter, we report on the first numerical simulations of the evolution of a scalar cloud initially bound to a BH. We restrict our analysis to the non-relativistic regime, since superradiant dark matter production was shown to be efficient for $\alpha\ll1$ and, moreover, taking into account that the the mass coupling is further reduced in the final stages of the PBH evaporation. 

Given the numerical challenges involved, we consider only spherically symmetric clouds and neglect the sub-leading effects of the BH spin. Note that spin-1 superradiant clouds have a spherical topology and that, as mentioned above, stimulated capture also populates mainly the ``1s''  state, so our results are directly applicable to these cases. We will nevertheless argue that the main features of the evolution should also apply to superradiant scalar clouds, which only have axial symmetry. 

In the non-relativistic regime $\alpha\ll 1$, the Klein-Gordon equation in a curved space-time reduces to the Schr\"odinger equation with a gravitational potential. The latter includes contributions from the BH, $V_{BH}\simeq -\alpha/r$, and from the scalar cloud's self-gravity, $V_{c}$. These can be added linearly in this regime, given the large hierarchy between the BH and the gravitational Bohr radius, even for a scalar cloud mass $M_c\gtrsim M_{BH}$. In this regime, Einstein's field equations reduce to the Poisson equation, which therefore determines the self-gravity potential:
\begin{eqnarray} \label{SP}
i\dot\psi &=& -{1\over 2\mu}\nabla^2\psi+(V_{BH}+V_C)\psi~,\nonumber\\
\nabla^2V_c &=&\mu^2 G|\psi|^2~,
\end{eqnarray}
where the wavefunction is normalized such that $\mu\int d^3r |\psi|^2=M_c$, i.e.~$|\psi|^2$ is the particle number density. Since the cloud's backreaction is small close to the event horizon, we may assume that the BH evaporates according to Hawking's semi-classical result \cite{Hawking:1975vcx}:
\begin{eqnarray}
M_{BH}(t) = M_{BH,i}\left(1-t/\tau\right)^{1/3}
\end{eqnarray}
where $\tau=M_{BH,i}^3/3e_TM_P^4$ is the lifetime of a BH with initial mass $M_{BH,i}$. The coefficient $e_T$ depends on the particles emitted by the BH and, hence, on its Hawking temperature. For light PBHs with $M_{BH,i}\lesssim 10^6$ kg, this is above the TeV scale, which means that all Standard Model particles are emitted and $e_T\simeq 4.4\times 10^{-3}$ \cite{March-Russell:2022zll}, up to a small contribution from dark matter. While new exotic particles may also contribute to $e_T$, this is irrelevant to our analysis, where we take $\tau$ as a free parameter.

If the scalar cloud is formed for $t\ll \tau$ with a mass $M_c\ll M_{BH}$, the above system reduces to a Hydrogen-like Schr\"odinger problem with a Coulomb-like potential, thus yielding bound states with the spectrum given in Eq.~(\ref{H_spectrum}). However, as the BH evaporates and its mass decreases, for $t\lesssim \tau$, the scalar field's self-gravity becomes significant, and we need to solve the full non-linear Schr\"odinger-Poisson system above with a time-dependent Hamiltonian.

For numerical purposes, it is useful to express the above differential equations in terms of dimensionless variables. Defining the cloud's mass coupling $\alpha_c=\mu M_c/M_P^2\ll 1$, we consider the dimensionless time and radial variables $\hat{t}= 2\alpha_c^2\mu t$ and $\hat{r}=2\alpha_c\mu r$, respectively, and express the potential energies as $\hat{V}=V/(2\alpha_c^2\mu)$, both for the BH and cloud contributions. This means that we normalize all quantities to the typical scales of a self-bound scalar cloud (i.e.~boson star). We also rescale the wavefunction such that $\int d\hat{r}~ \hat{r}^2|\hat{\psi}|^2=1$, obtaining the system of coupled differential equations:
\begin{eqnarray} \label{SP_rescaled}
&i\partial_{\hat{t}}\hat{\psi} = -{1\over \hat{r}}{d^2\over d\hat{r}^2}(\hat{r}\hat{\psi})+\left(-{M_{BH}/M_c\over \hat{r}}+\hat{V}_c\right)\hat{\psi}~,\nonumber\\
&{1\over \hat{r}}{d^2\over d\hat{r}^2}(\hat{r}\hat{V}_c) =|\hat{\psi}|^2~.
\end{eqnarray}
This rescaling also shows that the problem depends only on two parameters, the initial mass ratio $M_{BH,i}/M_c$ (typically $\gtrsim \alpha^{-1}\gg 1$ for superradiant clouds) and the rescaled BH lifetime $\hat{\tau}=2\alpha_c^2\mu\tau = (2\alpha_c^2\alpha/3e_T)(M_{BH,i}/M_P)^2\gg 1$ for $M_{BH,i}\gg M_P$. This poses a considerable numerical challenge, requiring long simulations with a large spatial grid to follow the evolution of an initially small cloud bound to the BH as it expands by a few orders of magnitude and potentially becomes (partially) unbound. 

In practice, we can only study the system's evolution starting with a BH mass that is already only a few times the scalar cloud's mass. However, as argued above, we expect the initial evolution to occur in the adiabatic regime, where $\psi$ adjusts to the slowly varying potential and follows the stationary solutions with decreasing BH mass. This implies that an accurate description can be obtained if we use the latter as initial conditions. 

Adiabaticity can be quantified in terms of the relative change in the wavefunction's frequency (or equivalently its binding energy) within an oscillation period, i.e.~the evolution is adiabatic if
\begin{eqnarray}
\xi={\dot\omega_B\over \omega_B^2}\simeq {8\over 3\hat{\tau}}\left({M_c\over M_{BH,i}}\right)^2\left(1-{\hat{t}\over\hat{\tau}}\right)^{-5/3}\ll 1
\end{eqnarray}
where in the last expression we used the binding energy of the ``1s'' Hydrogen-like bound state corresponding to the initial BH mass. While $\omega_B$ includes contributions from the cloud's self-gravity, these only increase it, so that this parameter is a good indicator of the adiabaticity of the evolution. It is clear from this expression that the evolution should become non-adiabatic close to the BH evaporation time, but initially $\xi\ll 1$ even for $M_{BH,i}/M_c\sim$ few given that typically $\hat{\tau}\gg 1$.

We therefore consider initial conditions corresponding to a stationary solution of Eqs.~(\ref{SP_rescaled}), $i\partial_{\hat{t}}\hat\psi=\hat\omega\psi$. To obtain this solution for a given value of the ratio $M_{BH,i}/M_c$, we employ a semi-numerical iterative procedure as follows. We start by replacing the ground state wavefunction for the pure Coulomb problem (i.e.~neglecting the cloud's self-gravity), $\hat{\psi}^{(0)}\propto e^{-{M_{BH,i}\over 2M_c}\hat{r}}$, on the right-hand side of the Poisson equation in \eqref{SP_rescaled}, solving this to obtain the leading self-gravity potential, $\hat{V}_c^{(0)}$. We then substitute this in the Schr\"odinger equation and use linear perturbation theory ($M_c< M_{BH,i}$) to compute the first-order correction to the bound state wavefunction, $\hat{\psi}^{(1)}$. We use this corrected wavefunction as an initial guess for a self-consistent method to numerically find the stationary solution of the Schr\"odinger-Poisson system, until the wavefunction and the corresponding binding energy $\hat{\omega}$ converge within a prescribed accuracy.

For the numerical evolution of the Schr\"odinger-Poisson system we employ a modified Crank-Nicolson method with a predictor scheme described in the Suplemental Material appended to this Letter. This type of method is particularly appropriate to solve Schr\"odinger-like differential equations since, by construction, it preserves the wavefunction norm and average energy (which we may use to quantify numerical error). We performed numerical simulations for different values of the mass ratio $M_{BH,i}/M_c$ and the BH's rescaled lifetime $\hat{\tau}$, using the Adamastor cluster at the Center for Physics of the University of Coimbra. Our numerical code and animations for the simulations labeled A, B and C, with $M_{BH,i}/M_c=5$ (see Fig.~\ref{energy_simulations} below), are publicly available in \cite{github}.

In Fig.~\ref{simulation} we show the evolution of the particle number density in simulation A ($\hat{\tau}=1.5\times10^5$), where the majority of the cloud remains self-bound. As expected, while in the adiabatic regime $\xi<1$, the wavefunction tracks the instantaneous stationary solutions of the Schr\"odinger-Poisson system for the corresponding BH mass (dashed gray curves), which we obtained using the iterative procedure described above. This is also observed in all other simulations. 

\begin{figure}[h!]
    \centering
    \includegraphics[width=\linewidth ]{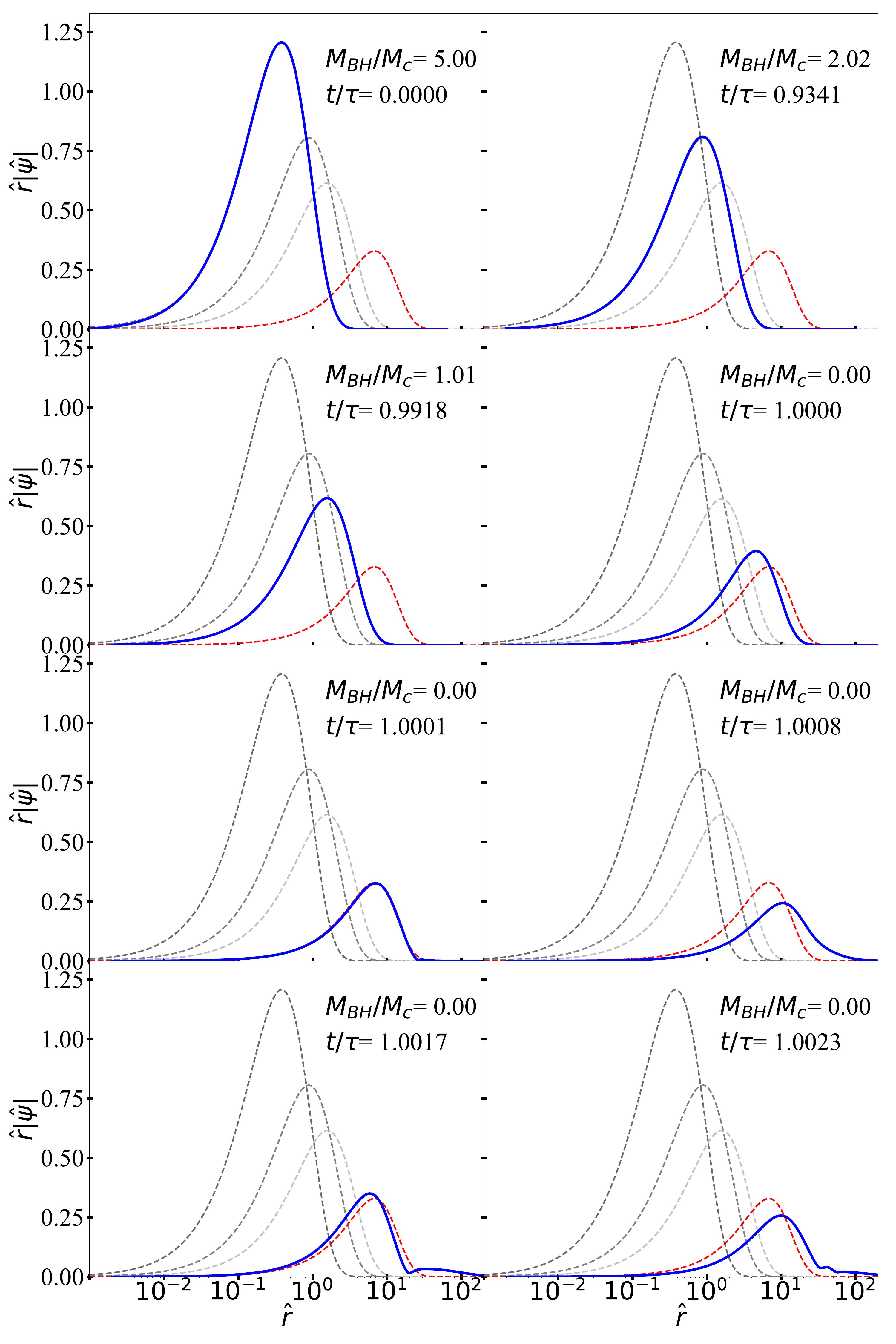}
    \caption{Numerical evolution of the (square-root of the) particle number density (blue) for simulation A with $M_{BH,i}/M_c=5$ and $\hat\tau=1.5\times 10^5$. Also shown are the stationary solutions for $M_{BH}/M_c=5,2,1$ (dashed gray) and the boson star ground state solution (dashed red).}
    \label{simulation}
\end{figure}

In this example, the evolution becomes non-adiabatic ($\xi=1$) when $M_{BH*}\simeq 0.3 M_c$, after which $|\hat{\psi}|$ oscillates (roughly) about the boson star solution with mass $M_c$, obtained numerically as in \cite{Ruffini:1969qy}. One also identifies a small tail moving towards larger distances. These results suggest that most of the scalar cloud remains self-bound in this example, and likely in a superposition of two states.

\begin{figure}[h!]
    \centering
    \includegraphics[width=\linewidth ]{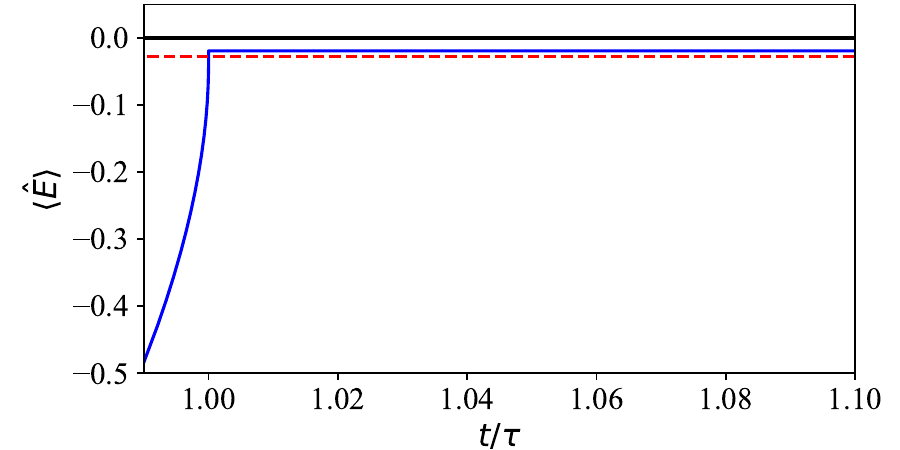}
    \caption{Evolution of the energy expectation value for $M_{BH,i}=5M_c$ and $\hat\tau=1.5\times 10^5$ (blue), compared with the energy of the boson star solution with the same mass (dashed red).}
    \label{energy}
\end{figure}

To better assess the results of this and other simulations, we have computed the (rescaled) energy expectation value:
\begin{equation} \label{average_energy}
\langle \hat{E}\rangle = \int d\hat{r}\,\hat{r}^2\left[|\partial_{\hat{r}}\hat{\psi}|^2+\left(\hat{V}_{BH}+{1\over2}\hat{V}_c\right)|\hat\psi|^2\right]~,     
\end{equation}
 which is conserved, from Eqs.~(\ref{SP_rescaled}), up to the BH mass decrease. Since $\langle \hat{E}\rangle<0$ only if the final state is at least partially bound, we may use it as a diagnostic tool.
 In Fig.~\ref{energy} we show its evolution in simulation A, for which $\langle \hat{E}\rangle$ increases as the BH evaporates but stabilizes at a negative value that is only $\simeq 30\%$ larger than the boson star energy $E_{BS}$ (dashed red line). This may be attributed to both the mass loss and self-bound state superposition.

To quantify the mass loss, we computed the evolution of the probability of finding particles at distances $\hat{r}>40$ (i.e.~well beyond the boson star radius), finding that it asymptotes to a constant value of $\simeq 7\%$. This means that $\simeq 93\%$ of the cloud remains self-bound after the BH evaporates completely in this example.

Most importantly, our simulations revealed that the value of $\langle \hat{E}\rangle$ after the BH evaporates depends only on the BH-to-scalar cloud mass ratio at the non-adiabatic point ($\xi=1$), $M_{BH,*}/M_c=(8/3\hat\tau)^{1/5}(M_{BH,i}/M_c)^{3/5}$, as shown in Fig.~\ref{energy_simulations}. This conclusion is supported by additional simulations performed with $M_{BH,i}/M_c=2$, for the same values of $M_{BH,*}/M_c$ as in simulations A$-$C, yielding essentially the same scalar cloud evolution.

\begin{figure}[h!]
    \centering
    \includegraphics[width=\linewidth ]{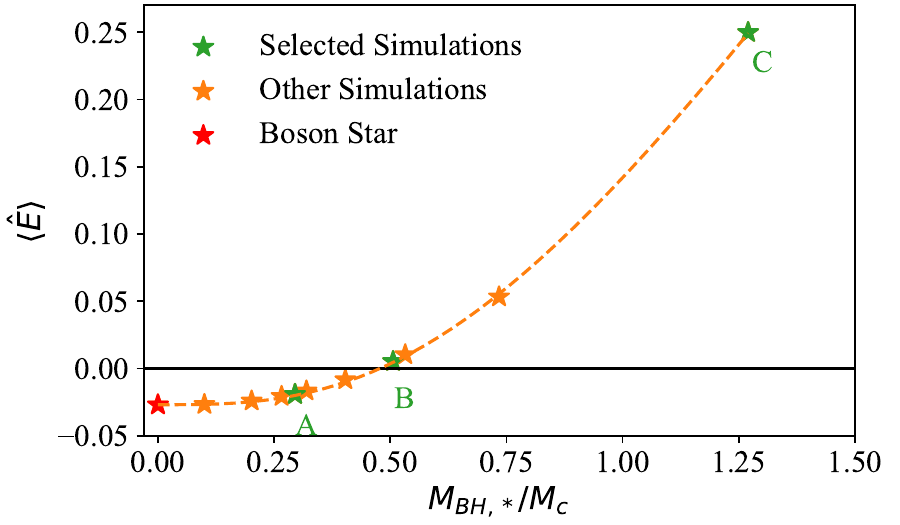}
    \caption{Final energy expectation value as a function of the BH-to-cloud mass ratio at the non-adiabatic point.}
    \label{energy_simulations}
\end{figure}

In particular, we find that $\langle\hat{E}\rangle<0$ for $M_{BH,*}/M_c<0.47$, so that the cloud remains at least partially self-bound if at the non-adiabatic point (or ``quench'') the BH's mass is already less than one half of the cloud's mass, in line with our qualitative expectation. As shown in Fig.~\ref{energy_simulations}, the boson star energy is attained in the limit $M_{BH,*}/M_c\rightarrow 0$, which would correspond to a fully adiabatic transition from scalar cloud to boson star.

 In addition, our results also suggest that a fraction of the cloud may remain self-bound even if the average energy is positive after the BH evaporates, provided that $\langle\hat{E\rangle}\lesssim |E_{BS}|$. For instance, in simulation B ($\hat{\tau}=10^4$), we also observe an oscillation of the wavefunction's norm close to the boson star solution, although with a smaller peak value and a more pronounced mass loss ($\simeq 33\%$) than in simulation A.

In simulation C ($\hat{\tau}=10^2$), as well as for the other simulations with $M_{BH,*}> 0.5M_c$ and $\langle\hat{E\rangle}\gtrsim |E_{BS}|$, we observe that the wavefunction's peak  moves towards larger distances throughout the simulation, showing that in this regime the scalar cloud dissipates away. 

We may thus robustly conclude that  a fraction of the scalar cloud remains self-bound as a boson star for $M_{BH,*}/M_c\lesssim 0.5$. Although the resulting superposition of bound states and its stability merit further investigation in the future, this result is already remarkable, yielding a new mechanism for boson star formation seeded by scalar clouds around evaporating BHs, as conjectured in \cite{March-Russell:2022zll}. 

Although our results do not strictly apply to scalar clouds from BH superradiance, we expect the adiabaticity arguments confirmed by our numerical simulations to be independent of the BH and cloud's angular momentum. Hence, we also expect superradiant clouds to become spinning boson stars after their host BH evaporates away, provided the evolution remains adiabatic until the BH's becomes somewhat lighter than the cloud (noting that spinning boson stars are unstable and decay into the non-spinning ground state \cite{Sanchis-Gual:2019ljs}). A more quantitative statement can, of course, only be made with dedicated numerical simulations including the angular dependence of the cloud's wavefunction and the effects of the BH spin, which we also plan to perform in the future.

While deviations from Hawking's semi-classical BH evaporation law could in principle affect our conclusions, they will only be significant before the non-adiabatic point. The formation e.g.~of stable Planckian remnants is therefore irrelevant for $M_c\gg M_P$. Furthermore, any effect that slows down the BH evaporation, such as potentially the recently conjectured {\it memory burden} effect \cite{Dvali:2020wft}, should make the scalar cloud's evolution adiabatic until later stages, therefore promoting boson star formation.

Our conclusions are particularly relevant for dark matter production through primordial BH superradiance, since \cite{March-Russell:2022zll} identified parametric regions where the resulting dark matter clouds satisfy the condition $M_{BH,*}/M_c\lesssim 0.5$. This strongly suggests that dark matter can be produced via a purely gravitational mechanism, with no need for other interactions with Standard Model particles, and that a significant fraction of dark matter may presently exist in the form of self-bound microscopic boson stars with very large occupation numbers. This means that, even if individual dark matter particles interact too weakly with known particles, these dark boson stars may be detectable due to e.g.~coherent enhancement of scattering cross sections \cite{Hardy:2015boa}. We note that even if only a small fraction of the scalar cloud remains self-bound after the BH evaporates, the resulting boson star occupation numbers may still be quite large. The detectability of such boson stars depends on the nature and strength of the interactions between dark matter and Standard Model particles, and we hope that our results motivate future research in this topic.
\vfill

\section*{Acknowledgments}
We thank Miguel Zilh\~ao for useful discussions. D.~N.~is supported by the doctoral grant 2024.02437.BD funded by FCT - Funda\c{c}\~ao para a Ci\^encia e Tecnologia, I.P. This work was supported by national funds by FCT, through the research projects with DOI identifiers 10.54499/UID/04564/2025, 10.54499/CERN/FIS-PAR/0027/2021, and by the projects 10.54499/2024.00252.CERN and 10.54499/2024.00249.CERN funded by measure RE-C06-i06.m02 – ``Reinforcement of funding for International Partnerships in Science, Technology and Innovation'' of the Recovery and Resilience Plan - RRP, within the framework of the financing contract signed between the Recover Portugal Mission Structure (EMRP) and the Foundation for Science and Technology I.P. (FCT), as an intermediate beneficiary.

\begin{appendix}

\section*{Supplemental Material}

Here we give some technical details on the numerical method employed in our simulations, adapted from the modified Crank-Nicolson method with a predictor scheme proposed in \cite{Ringhofer}, where we completed the evolution with a self-consistency method.

 We work in a 1-dimensional radial grid with $N$ points labeled $x_n$,  $n=1,\ldots, N $, with $\Delta x_n = x_{n+1}-x_{n}$. The time evolution is also performed in discrete steps with $\Delta t_k= t_{k+1}-t_k$. For any function of the space and time coordinates, we use the notation $f_n^k\equiv f(x_n,t_k)$. Spatial derivatives are computed numerically using the standard finite-differences prescription, and denoted as $D_if_n^k$, where $i$ is the order of the derivative. The time derivative is defined as $\delta_t f^k = (f^{k+1}-f^{k})/\Delta t _k$ and we also define the time average as $\mu_t f^k = (f^{k+1} +f^{k})/2$.

It is also convenient for the numerical implementation to work with the variables $\hat{u}=\hat{r}\hat{\psi}$ and $\hat{v}=\hat{r}\hat{V}_c$ for the self-gravity potential. To numerically preserve the wavefunction norm and the conserved energy (for constant BH mass), we employ the following discretization scheme to the system in Eq. (\ref{SP_rescaled}) \cite{Ringhofer}:
\begin{eqnarray}
	\label{eq: system k}
		&{}& i \frac{u_n^{k+1 }-u_n^k}{\Delta t_k} +\mu_t D_2 u^k_n + \mu_t \left(\frac{v^k_n}{r_n}+V^k_{BH, n}\right) \mu_t u^k_n = 0 \nonumber\\
        &{}&D_2 v^k_n = -\frac{|u^k_n|^2}{r_n}~,
\end{eqnarray}
where, for simplicity, we have dropped the hatted notation, although we are implicitly using rescaled variables.
This system cannot be solved directly, given that the solutions $u^{k+1}$ and $u^k $ are coupled. For this we use a predictor scheme along the lines proposed in \cite{Ringhofer} following the time evolution chain $u^k \rightarrow u^{k,1}\rightarrow u^{k,2}\rightarrow u^{k+1}$ with two intermediate steps, where we find approximate solutions. We first define $\delta_t ^{(i)} f^k = (f^{k,i}-f^{k})/\Delta t _k $ and  $\mu_t ^{(i)} f^k  = (f^{k,i}+f^{k})/2$, where $i=1,2$ labels the intermediate steps. The predictor scheme starts by computing $u^{k,1}$ and then $v^{k,1}$ with the cloud potential at $t_k$ as follows:
\begin{equation}
	\label{eq : system k1}
	\begin{aligned}
		&{} i \frac{u_n^{k,1 }-u_n^k}{\Delta t_k} +\mu_tD_2 u^k_n +  \left(\frac{v^k_n}{r_n}+V^k_{BH, n}\right) \mu^{(1)}_t u^k_n = 0 \\ &
		D_2 v^{k,1}_n = -\frac{|u^{k,1}_n|^2}{r_n}~.
	\end{aligned}
\end{equation}
Since $u^{k,1}$ and $u^{k}$ are now decoupled, the solution after the first intermediate step is given, in matrix form, by:
\begin{equation}
	\label{eq : system k1 matrix}
		{} \textbf{u}^{k,1}= \textbf{A}_+^{-1}\textbf{A}_- \textbf{u}^k~, \qquad
		\textbf{v}^{k,1} = \textbf{D}^{-1}_2 \left(-\frac{|\textbf{u}^{k,1}|^2}{\textbf{r}}\right)~,
\end{equation}
where $\textbf{u}$ and $\textbf{v}$ are vectors, e.g.~$\textbf{u}^k \equiv (u^k_1, u^k_2,...,u^k_N) $, and  $\textbf{A}_\pm = \frac{i}{\Delta t_k}\textbf{I} \pm \frac{1}{2} \textbf{D}_2 \pm \textbf{diag} \left(\frac{v^k}{2 r} + \frac{V^k_{BH}}{2} \right)$ are $N\times N$ matrices, where $\textbf{I}$ is the identity matrix.  $\textbf{D}_2$ and $\textbf{D}^{-1}_2$ are the numerical $N\times N$ Laplacian matrix and its inverse.

The solution in the second intermediate step, $u^{k,2}$, is then obtained by replacing $u^{k,1}\rightarrow u^{k,2}$  and $v^{k} \rightarrow \mu_t ^{(1)}v^{k} $ in Eq.~(\ref{eq : system k1}), yielding:
\begin{equation}
	\label{eq : system k2}
		\textbf{u}^{k,2}= \textbf{B}_+^{-1}\textbf{B}_- \textbf{u}^k~, \qquad
		\textbf{v}^{k,2} = \textbf{D}^{-1}_2 \left(-\frac{|\textbf{u}^{k,2}|^2}{\textbf{r}}\right)~,
\end{equation}
where $\textbf{B}_\pm = \frac{i}{\Delta t_k}\textbf{I} \pm \frac{1}{2} \textbf{D}_2 \pm \textbf{diag} (\frac{v^{k,1}+v^k}{2 r} + \frac{\hat{V}^{k+1}_{BH} }{2} )$.

Finally, the energy is numerically preserved by taking   $v^{k+1} = v^{k,2}$ and $u_n^{k+1} = e^{i\alpha_n^k} \, u_n^{k,2}$. To obtain the phase we may write it as $\alpha_n^k =  i \Delta t_k^3 \eta g_n$, where $g$ is some real bounded function, and $\eta$ is a real parameter that can be obtained via \cite{Ringhofer}:
\begin{equation}
	\eta=  \frac{ (\textbf{v}^{k,1}/\textbf{r}-\textbf{v}^{k,2}/\textbf{r})^\textbf{T}  (\delta_t^{(2)} |\textbf{u}|^2 ) }{2\text{Re}\left[ i D(\textbf{g} |\textbf{u}^{k,2}|) ^\textbf{T}  D |\textbf{u}^{k,2}|  \right] } .
\end{equation}
Although this parameterization already provides a very good approximation, we may further improve the $u^{k+1}$ solution by performing a self-consistent cycle using the system in Eq.~(\ref{eq: system k}). We take the potential $v^{k,2}$ as a guess in order to compute a new $u^{k+1}$ solution, then we recalculate the self-gravity potential and repeat the procedure until it converges within a prescribed accuracy. 

\end{appendix}

\end{document}